\documentclass[a4paper,11pt]{article}
\usepackage{geometry,setspace}                
\usepackage{amsmath,amssymb,amsfonts,dcolumn,color,graphicx,graphics,latexsym,placeins,epsfig,caption}
\usepackage{authblk}
\usepackage{subcaption}
\usepackage{rotating,bm,mathrsfs}
\usepackage{cite}
\usepackage[title]{appendix}

\usepackage[pagebackref=false, colorlinks=true]{hyperref}
\definecolor{redish}{rgb}{0.7,0.2,0.0}  
\definecolor{bluish}{rgb}{0.2,0.5,0.8}

\hypersetup{linkcolor=redish,       
            citecolor=bluish,       
            filecolor=magenta,      
            urlcolor=bluish}        

\begin{document}

\date{}
\author{Suvankar Paul \thanks{\href{mailto:suvankar266@gmail.com}{suvankar266@gmail.com}}}
\affil{\it Department of Physics, Raiganj Surendranath Mahavidyalaya, Raiganj 733134, India}

\title{Strong Gravitational Lensing by Compact Object without Cauchy Horizons in Effective Quantum Gravity}

\maketitle

\begin{abstract}
{\small In this work, we theoretically investigate strong gravitational lensing effects and evaluate various lensing observables of a static, spherically symmetric solution in the context of effective quantum gravity (EQG). Among the three types of solutions proposed in EQG backgrounds, this is the third type without having Cauchy horizons. This solution gives rise to black hole as well as horizonless wormhole solutions depending on the range of values of the parameters of the theory. Based on the data from SgrA* and M87* observations, possible bounds on the parameter are obtained. It is found that the horizonless wormhole solution is ruled out by SgrA* observations, but is allowed by M87* observations. We analyze and provide estimates of the lensing observables, some of which can potentially be detected by observational tools.}
\end{abstract}

\onehalfspacing

\section{Introduction}
\label{sec:intro}
General Relativity (GR) \cite{Einstein}, as a classical theory of gravity, has been remarkably successful in describing gravitational effects on macroscopic scales in weak field regime, like our solar system. However, it faces some fundamental challenges in strong field regime, such as the presence of singularities and its incompatibility with quantum physics \cite{Penrose,Hawking,Ashtekar1}. This has spurred extensive research on various modified theories of gravity, including elusive quantum theories of gravity \cite{Clifton,Mukohyama,Rovelli,Thiemann}. Although it is believed that a full quantum theory of gravity should resolve the singularity problem, in the absence of it, various phenomenological approaches have been proposed in the literature to resolve this issue. Some of these include generating singularity-free regular black hole solutions \cite{Bardeen, Hayward, Simpson-Visser,Maeda,Lan}, considering non-commutative geometry \cite{Nicolini,Nasseri}, using frameworks like Loop Quantum Gravity (LQG) \cite{Ashtekar2,Perez,Bojowald,Bohmer,Chiou,Saini,Bodendorfer}, etc. A similar approach to incorporate quantum corrections to classical black hole singularities is to consider general covariance in semiclassical effective models of quantum gravity described by Hamiltonian constraints. Recently, within such an Effective Quantum Gravity (EQG) framework, two distinct static solutions were first derived in Ref. \cite{Zhang1}, and later a third one was obtained in Ref. \cite{Zhang2} by Zhang \emph{et al}. The first two types of solutions of Ref. \cite{Zhang1} have spurred a plethora of studies in black hole physics, such as quasinormal modes \cite{Konoplya24,Malik}, gravitational lensing \cite{Liu1,Wang}, shadows \cite{Liu2}, accretion disk imaging \cite{Shu} and cosmic censorship conjecture \cite{Lin}. In addition, similar studies have also come up on the rotating version of the solutions in Ref. \cite{Ban}, applying modified Newman-Janis algorithm \cite{Azreg1,Azreg2}. However, the third type of solution in Ref. \cite{Zhang2} remains relatively unexplored (apart from a recent study on light rings and shadows in Ref. \cite{Liu3}), even though it is quite interesting to admit both black hole and wormhole solutions. In this context, we carry out investigations on this interesting solution with regard to its strong lensing characteristics and compare our results with observational data. This establishes the motivation for the present work.

Ever since the very first expedition to observe deflection of light by the Sun during a solar eclipse in 1919 \cite{Eddington}, gravitational lensing has become a pioneering tool to test GR. It has been remarkably successful in the weak gravity regime, for example, around celestial bodies in our Solar system. 
However, in light of the recent imaging of black holes of M87* \cite{EHTM1,EHTM5,EHTM6} and SgrA* \cite{EHTS1,EHTS4,EHTS6} by the Event Horizon Telescope\footnote{\href{https://eventhorizontelescope.org/}{https://eventhorizontelescope.org/}} (EHT) collaboration, a new era of high precision observational astronomy to dig deeper into the strong gravity regime has opened up. Since then, studies of strong gravitational lensing, shadows, accretion disk images etc. have become ubiquitous to estimate various parameters associated with such compact objects. It is believed that centers of galaxies are inhibited by supermassive black holes, making them the best candidates to probe strong gravity phenomena. Accordingly, a substantive amount of research about strong gravitational lensing in the backgrounds of black holes have enriched the literature \cite{Virbhadra00,Frittelli,Bozza01,Bozza02,Eiroa,Hsieh,Ghosh,Ding} (for a comprehensive review, see also Ref. \cite{BozzaReview}).

However, it should be noted that it is the presence of photon sphere that usually governs the strong lensing characteristics of compact objects. A photon sphere (in spherically symmetric solutions) consists of unstable light rings such that a small perturbation can cause them to either get trapped inside the spherical surface without ever coming out of it, or to move towards asymptotic infinity. It was first formulated by Bozza in Ref. \cite{Bozza02} and later refined by Tsukamoto in Ref. \cite{Tsu17} that the deflection angle of light diverges logarithmically in the vicinity of a photon sphere, and corresponding analytic expressions of bending angle parameters were derived. The presence of such photon spheres is not limited to black holes only. Horizonless compact objects too can posses photon spheres. As a result, they may exhibit identical strong lensing features as that by black holes, making them behave like black hole mimickers. Therefore, horizonless compact objects have also attracted much attention in recent times \cite{UCO1,UCO2,UCO3,Vagnozzi}. Accordingly, various aspects of strong lensing by horizonless objects such as wormholes, naked singularities, gravastars, etc. have been studied in the literature \cite{WL4,WL5,WL6,WL7,WL11,WL12,WL13,WL15,WL19,WL22,WL23,NS2,NS3,NS4,NS5,gravastar,Cunha17,Cunha18,Hod18}. Another line of study has appeared in Refs. \cite{Patil,Shaikh19} where strong lensing features due to the presence of anti-photon sphere (stable light rings) have been analyzed, and a new analytic formulation of logarithmic divergence of bending angle due to anti-photon sphere has been derived. Moreover, a recent work in Ref. \cite{Paul} has uncovered an intriguing example of strong lensing without having a photon or anti-photon sphere by a null naked singularity. Interestingly, it turns out that the corresponding divergence of bending angle is non-logarithmic in nature (see Ref. \cite{Tsu20} also, for another case of non-logarithmic divergence).
A general overview of these studies is that, while strong lensing features of black holes and horizonless objects may sometimes become similar, they usually exhibit dramatically different characteristics.

In this paper, we extend our research along similar line with an in depth study of strong lensing, considering both black hole and wormhole cases in the context of the third type of solution of Ref. \cite{Zhang2} and exemplify our analysis with SgrA* and M87* observations. This paper is organized as per the following outline:
section \ref{sec:2} starts with a brief recapitulation of the nature and properties of the spacetime of Ref. \cite{Zhang2}. Then in Sec. \ref{sec:3}, we discuss the geodesic structure of light rays and obtain the location of photon sphere from the corresponding effective potential. Next, in Sec. \ref{sec:4}, we put forward possible bounds on one of the free parameters of the theory using observational data of SgrA* and  M87*. Section \ref{sec:5} deals with the study of light deflection in the strong field regime. After that, we compute various lensing observables and discuss upon possible scenarios for futuristic observation in Sec. \ref{sec:6}. Finally, the paper is concluded with a brief summary of the results and discussion on possible future research prospects in Sec. \ref{sec:7}.\\
Unless specifically mentioned, we have considered $c=G=1$ units throughout.


\section{The Metric and its Properties}
\label{sec:2}
\begin{figure}[h]
\centering
\includegraphics[scale=1.2]{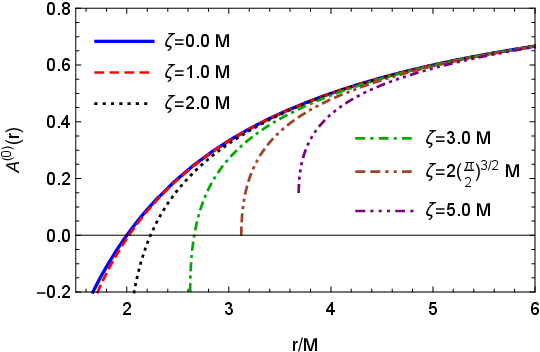}
\caption{Plot of $A^{(0)}(r)$ as a function of $r$ (in units of $M$) for different values of $\zeta$. For $\zeta/M = 0.0,1.0,2.0,3.0$ \& $2\left(\frac{\pi}{2}\right)^{3/2}$ event horizons do exist, with $\zeta/M=2\left(\frac{\pi}{2}\right)^{3/2}$ representing the extremal black hole. However, for $\zeta/M=5.0$, event horizon does not exist.}
\label{fig:Aplot}
\end{figure}

\begin{figure}[h]
\centering
\includegraphics[scale=0.9]{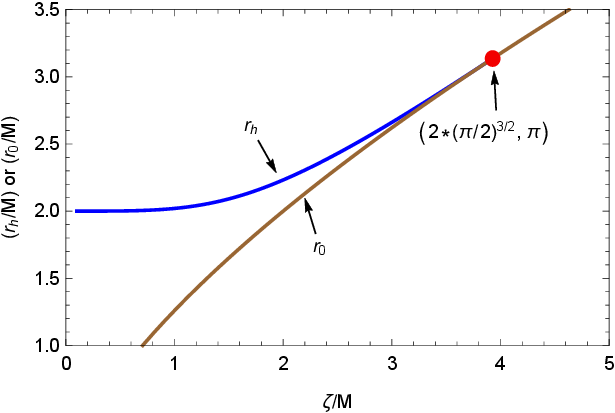}
\caption{Combined plot of $r_h$ and $r_0$ (in units of $M$) versus $\zeta$ (in units $M$). It can be seen that $r_h>r_0$ for $\zeta/M<2\left(\frac{\pi}{2}\right)^{3/2}$ representing a black hole. The extremal black hole case is denoted by the red dot having coordinates $\left(2\left(\frac{\pi}{2}\right)^{3/2},\pi\right)$. Beyond this point, $r_h$ does not exist signifying a wormhole spacetime with $r_0$ being the throat.}
\label{fig:Rhplt}
\end{figure}

The issue of general covariance in the Hamiltonian framework of semiclassical effective theories of gravity, resulting from canonical quantum gravity models, has been long standing. Recently, this issue is addressed in case of spherically symmetric, vacuum gravity by Zhang {\em et al.} in Refs. (\cite{Zhang1}, \cite{Zhang2}). Two families of quantum-modified effective spacetimes were first proposed in Ref. \cite{Zhang1}, and a third one, without introducing Cauchy horizons, has been subsequently obtained in Ref. \cite{Zhang2}. For a brief review of the formalism of Ref. \cite{Zhang2}, see Appendix \ref{appen}. Each of the three quantum-corrected solutions, thus obtained, is parameterized by its own free parameters that account for the quantum gravity effects. Setting them to zero returns the classical black hole solutions. In this paper, we discuss the third solution without Cauchy horizons, given by the following static, spherically symmetric line element
\begin{equation}
ds^{2}=-A^{(q)}(r)dt^{2}+\frac{dr^{2}}{\mu(r)A^{(q)}(r)}+C(r)(d\theta^{2}+\sin^{2}\theta d\phi^{2})
\label{eq:genmetric}
\end{equation}
where
\begin{equation}
A^{(q)}(r)=1-(-1)^q\frac{r^2}{\zeta^2}\arcsin\left(\frac{2M\zeta^2}{r^3}\right)-\frac{q\pi r^2}{\zeta^2}, ~~~ \mu(r)=1-\frac{4M^2\zeta^4}{r^6}, ~~~ \text{and} ~~~ C(r)=r^2.
\label{EQG1}
\end{equation}
Here, $M$ represents the mass and ($\zeta$, $q$) are the free parameters accounting for quantum gravity effects. For $q=0$, the metric function $A^{(0)}(r)$ takes the form
\begin{equation}
A^{(0)}(r)=1-\frac{r^2}{\zeta^2}\arcsin\left(\frac{2M\zeta^2}{r^3}\right).
\label{eq:A0}
\end{equation}
Naturally, setting the `quantum parameters' to zero (i.e., $q=0$ \& $\zeta\to0$) recovers the classical Schwarzschild geometry. The spacetime, for $q=0$, is asymptotically flat for any finite value of $\zeta\neq0$, so that $\zeta$ alone can be considered as the deviation parameter from the Schwarzschild geometry for $q=0$. In this case, $M$ becomes the ADM (Arnowitt–Deser–Misner) mass of the spacetime.\\
The $\arcsin$ function yields finite value if its argument remains within the range $[-1,1]$, which is satisfied for $r\ge (2M\zeta^2)^{1/3}$. This puts a lower limit on the value of $r$ with $r_{\text{min}}=(2M\zeta^2)^{1/3}$. From Eq. (\ref{eq:A0}), we see that $A^{(0)}(r)$ monotonically increases from a value $A^{(0)}(r_{\text{min}})=1-\frac{\pi}{2}\left(\frac{2M}{\zeta}\right)^{2/3}$ to a value equal to unity, as $r$ increases from $r_{\text{min}}$ to $\infty$. Plots of $A^{(0)}(r)$ as functions of $r$ (in units of $M$) for different values of $\zeta/M$ are shown in Fig. (\ref{fig:Aplot}).
\\
From the figure, it can be seen that $A^{(0)}(r)$ always remains positive, starting from a value $A^{(0)}(r_{\text{min}})>0$, when $\zeta/M>2\left(\frac{\pi}{2}\right)^{3/2}$. In this case, the equation $A^{(0)}(r)=0$ will never have any real solution and there will be no event horizon.
On the other hand, when $\zeta/M\leq2\left(\frac{\pi}{2}\right)^{3/2}$, $A^{(0)}(r_{\text{min}})\leq0$, so that the equation $A^{(0)}(r)=0$ yields one real positive root. As a result, the spacetime possesses an event horizon (black hole) when $\zeta/M\leq2\left(\frac{\pi}{2}\right)^{3/2}$ with the equality sign representing the extremal black hole case.\\
Let us now consider the function $\mu(r)=1-\frac{4M^2\zeta^4}{r^6}$ in the denominator of the $dr^2$-term of Eq. (\ref{eq:genmetric}). For $r\in[r_{\text{min}},\infty)$, $\mu(r)\in [0,1]$. Moreover, if the conditions $\mu(r)=0$ and $A^{(0)}(r)\neq0$ are simultaneously satisfied, then the metric component $g_{rr}$ will diverge, whereas $g_{tt}$ will remain finite. This particular situation resembles the throat of a wormhole. Therefore, we have two different kinds of radii -- $r_h$ representing the position of an event horizon (satisfying the condition $A^{(0)}(r)=0$, $\mu(r)\neq0$) and $r_0=r_\text{min}=(2M\zeta^2)^{1/3}$ corresponding to the location of a wormhole throat (satisfying the condition $\mu(r)=0$, $A^{(0)}(r)\neq0$). For the extremal case, $A^{(0)}(r_h)=\mu(r_0)=0$ where $r_h=r_0=M\pi$.\\
A combined plot of $r_h$ and $r_0$ (in units of $M$) versus $\zeta$ (in units $M$) is shown in Fig.(\ref{fig:Rhplt}), for convenience.
From this figure, it can be seen that $r_h>r_0$ for $\zeta/M<2\left(\frac{\pi}{2}\right)^{3/2}$, i.e., the event horizon is located outside the minimum radius or throat, representing a black hole. The extremal black hole case is denoted in the figure by the red dot having coordinates $\left(2\left(\frac{\pi}{2}\right)^{3/2},\pi\right)$. Beyond this point, i.e. $\zeta/M>2\left(\frac{\pi}{2}\right)^{3/2}$, the spacetime represents a wormhole with $r_0$ being the throat. To elaborate more on this aspect, let us consider the line element of a general static, spherically-symmetric wormhole of Morris-Thorne class as \cite{Morris} 
\begin{equation}
ds^2=-e^{2\Phi(r)}dt^2+\frac{dr^2}{1-\frac{\mathcal{B}(r)}{r}}+r^2\left(d\theta^2+\sin^2\theta d\phi^2\right)
\label{eq:MTgen}
\end{equation}
where $\Phi(r)$ is called the red-shift function and $\mathcal{B}(r)$ is known as the shape function. The wormhole throat ($r_0$) establishes the connection between two different regions and satisfies the condition $\left(1-\frac{\mathcal{B}(r)}{r}\right)|_{r_0}=0$, i.e., $\mathcal{B}(r_0)=r_0$. Additionally, $\mathcal{B}(r)$ satisfies the flare-out condition $\mathcal{B}'(r_0)<1$ \cite{Morris}. Moreover, $\Phi(r)$ must remain finite throughout (from the throat to infinity). If we compare the above line element with the line element of Eq. (\ref{eq:genmetric}) for $q=0$, we see that the function $\left(1-\frac{\mathcal{B}(r)}{r}\right)$ can be equated to the function $\mu(r)=1-\frac{4M^2\zeta^4}{r^6}$, as $\mu(r)|_{r_0}=0$. Therefore, we obtain $\mathcal{B}(r)=\frac{4M^2\zeta^4}{r^5}$, which implies that $\mathcal{B}'(r_0)=-5<1$. Hence, the flare-out condition is satisfied for this spacetime. Moreover, the $g_{tt}$-term of Eq. (\ref{eq:genmetric}), with $q=0$, also remains finite from $r_0$ to infinity. Therefore, it satisfies all the required conditions of a wormhole for $\zeta/M>2\left(\frac{\pi}{2}\right)^{3/2}$.\\
To summarize, it has been found that the spacetime given by Eq. (\ref{eq:genmetric}), with $q=0$, admits either a black hole or a wormhole corresponding to three possible scenarios as given below.
\begin{itemize}
\item Case-1: $\zeta/M<2\left(\frac{\pi}{2}\right)^{3/2}$ $\implies$ $r_h>r_0$ $\to$ Black hole
\item Case-2: $\zeta/M=2\left(\frac{\pi}{2}\right)^{3/2}$ $\implies$ $r_h=r_0=M\pi$ $\to$ Extremal Black hole
\item Case-3: $\zeta/M>2\left(\frac{\pi}{2}\right)^{3/2}$ $\implies$ $r_h$ does not exist $\to$ Wormhole
\end{itemize}
For $q>0$, the spacetime asymptotically resembles closely to the de Sitter geometry. In this case, the quantum parameters ($\zeta$, $q$) together play the analogous role of the cosmological constant in GR. If we assume a relationship like, $\Lambda=3q\pi/\zeta^2$, where $\Lambda$ is the effective cosmological constant, the metric function of Eq. (\ref{eq:genmetric}) can be written as
\begin{equation}
A^{(q)}(r)=1-(-1)^q\frac{r^2}{\zeta^2}\arcsin\left(\frac{2M\zeta^2}{r^3}\right)-\frac{\Lambda(\zeta,q)}{3}r^2.
\end{equation}
If the above function is expanded at large $r$, it can be approximated as
\begin{equation}
A^{(q)}(r)\approx 1-(-1)^q\frac{2M}{r}-\frac{\Lambda(\zeta,q)}{3}r^2.
\end{equation}
For even integer values of $q$, the metric function asymptotically resembles the Schwarzschild-de Sitter (SdS) spacetime. Whereas, for odd integer values of $q$, it produces gravitational effects analogous to that of a negative mass object \cite{Zhang2}. For a discussion on light rings and shadows for $q\neq0$, see Ref. \cite{Liu3}. Importantly, this spacetime does not contain any Cauchy horizon, which usually imply its stability under perturbations.


\section{Structure of Null Geodesics}
\label{sec:3}
We now consider the geodesic motion of photons in this spacetime. The line element of Eq. (\ref{eq:genmetric}) being spherically-symmetric, we can choose $\theta=\frac{\pi}{2}$ without any loss of generality. Therefore, the Lagrangian describing the motion of photon in this geometry can be written as
\begin{equation}
2\mathcal{L}=-A^{(q)}(r)\dot{t}^{2}+\frac{\dot{r}^{2}}{\mu(r)A^{(q)}(r)}+C(r)\dot{\phi}^{2}
\end{equation}
where an `overdot' represents a derivative with respect to the affine parameter along the geodesic. The above Lagrangian admits two constants of motion, namely
\begin{equation}
\frac{\partial\mathcal{L}}{\partial\dot{t}}=-A^{(q)}(r)\dot{t}=-E~,~
\frac{\partial\mathcal{L}}{\partial\dot{\phi}}=C(r)\dot{\phi}=L,
\end{equation}
where $E$ and $L$ represent the magnitudes of energy and angular momentum of a photon with respect to an asymptotic observer. From normalization of the four velocity vectors, $g_{\mu\nu}\dot{x}^\mu\dot{x}^\nu=0$, we obtain
\begin{equation}
\frac{\dot{r}^2}{\mu(r)}+V_{eff}(r)=E^2,  \hspace{0.5cm}
V_{eff}(r)=L^2\frac{A^{(q)}(r)}{C(r)},
\end{equation}
where $V_{eff}(r)$ is the effective potential corresponding to radial motion.\\
A light ray coming from a source at infinity may pass through a turning point at some radial distance $r_{\text{tp}}$ and then escapes to an asymptotically faraway observer. When the ray travels from infinity towards the turning point, $\dot{r}<0$, and when it moves away from the turning point, $\dot{r}>0$. Therefore, at the turning point, $\dot{r}|_{r_{\text{tp}}}=0$, or $V_{eff}(r_{\text{tp}})=E^2$. From this expression, we obtain a relation of the impact parameter $b$ $(=L/E)$ (which is a constant of motion) in terms of the turning point ($r_{\text{tp}}$) of light trajectory as, $b(r_{\text{tp}})=\sqrt{C(r_{\text{tp}})/A^{(q)}(r_{\text{tp}})}$.

\begin{figure}[h]
\centering
\begin{subfigure}{.5\textwidth}
\centering
\includegraphics[scale=0.93]{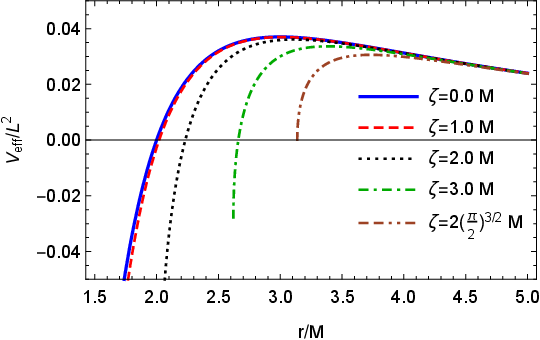}
\subcaption{Black hole}
\label{fig:Eff-potBH}
\end{subfigure}%
\begin{subfigure}{.5\textwidth}
\centering
\includegraphics[scale=0.57]{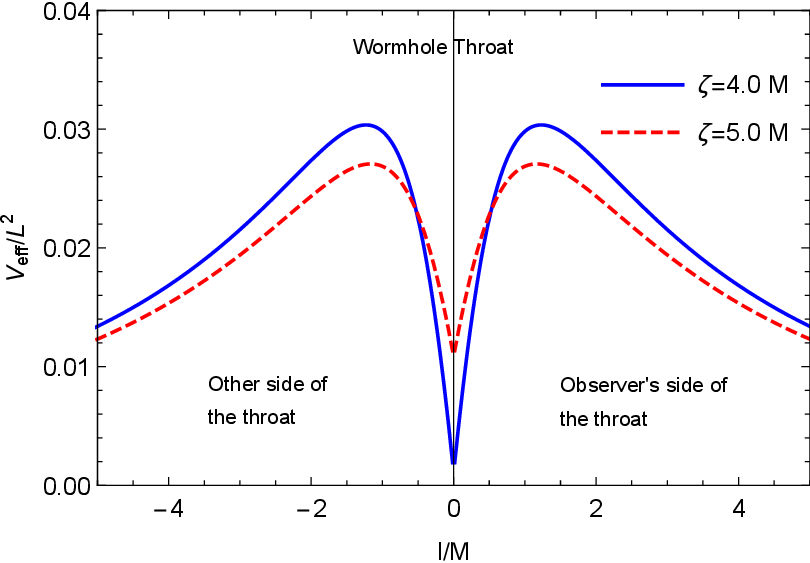}
\subcaption{Wormhole}
\label{fig:Eff-potWH}
\end{subfigure}
\caption{Plots of $V_{eff}/L^2$ as functions of (a) $r/M$ (Black hole) and (b) $l/M$ (Wormhole), with $q=0$, for different values of $\zeta$. The maxima of $V_{eff}$ correspond to the locations of photon spheres. In figure (a) for the black hole case, the radii of photon spheres increase with increasing value of $\zeta$. In figure (b) for the wormhole case, $V_{eff}$ is plotted against the proper radial coordinate $l$, such that at the throat $l(r_0)=0$.}
\label{fig:Eff-pot}
\end{figure}

In case of circular photon orbits, $\dot{r}=\ddot{r}=0$, which yield $V_{eff}=E^2$ and $dV_{eff}/dr=0$ respectively. The second condition implies that circular orbits correspond to the maxima or minima of the effective potential. In case of maxima ($d^2V_{eff}/dr^2<0$), the light rings are unstable which we generally call as photon spheres; and minima of the effective potential ($d^2V_{eff}/dr^2>0$) give rise to stable light rings, termed as anti-photon spheres \cite{Shaikh19}. Therefore, a photon sphere (or an anti-photon sphere) at $r=r_m$ corresponds to $V_{eff}|_{r_m}=E^2$ and $(dV_{eff}/dr)|_{r_m}=0$ which respectively result in
\begin{equation}
b_m=\sqrt{C(r_m)/A^{(q)}(r_m)} ~~~~ \text{and} ~~~~
\frac{C'(r_m)}{C(r_m)}-\frac{A^{(q)'}(r_m)}{A^{(q)}(r_m)}=0,
\label{eq:ps}
\end{equation}
where a prime denotes derivative with respect to $r$ and $b_m(=b|_{r_m})$ is called as the critical impact parameter corresponding to a photon sphere.\\
Putting the expressions of $A^{(q)}(r)$ and $C(r)$ in the second expression of Eq. (\ref{eq:ps}) and simplifying, we obtain the following equation for photon sphere,
\begin{equation}
r_m^6-9M^2r_m^4-4M^2\zeta^4=0.
\end{equation}
Interestingly, it is found that the above photon sphere equation does not depend on the parameter $q$. Therefore, the radius of the photon sphere will depend on the quantum parameter $\zeta$ only, as given below
\begin{equation}
r_m=M\sqrt{3+9\eta^{-1/3}+\eta^{1/3}} ~~~~ \text{where} ~~~~ \eta=27+2\tilde{\zeta}^4+2\tilde{\zeta}^2\sqrt{27+\tilde{\zeta}^4} ~~~ \text{and} ~~~ \tilde{\zeta}=\zeta/M.
\label{eq:rm}
\end{equation}
It can be easily verified from the above equation that $r_m=3M$ for $\zeta=0$, i.e., we recover the radius of the photon sphere in Schwarzschild spacetime.
The critical impact parameter ($b_m$) due to the photon sphere is obtained from the first expression of Eq. (\ref{eq:ps}) as given below
\begin{equation}
b_m=\left[\frac{1}{r_m^2}-\frac{q\pi}{\zeta^2}-\frac{(-1)^q}{\zeta^2}\arcsin\left(\frac{2M\zeta^2}{r_m^3}\right)\right]^{-1/2},
\label{eq:bm}
\end{equation}
where $r_m$ is given by Eq. (\ref{eq:rm}).\\
Plots of $V_{eff}/L^2$ as functions of (a) $r/M$ (Black hole) and (b) $l/M$ (Wormhole), with $q=0$, for different values of $\zeta$ (in units of $M$) are shown in Figs. (\ref{fig:Eff-potBH}) and (\ref{fig:Eff-potWH}) respectively. In Fig. (\ref{fig:Eff-potBH}) for the black hole case, the radii of photon spheres increase with increasing value of $\zeta$. Since minimum of the effective potential does not exist, there will no anti-photon sphere, which is usually the case for black holes. On the other hand, in Fig. (\ref{fig:Eff-potWH}) for the wormhole case, $V_{eff}$ is plotted against the proper radial coordinate $l$ which is defined as
\begin{equation}
l(r)=\pm\int_{r_0}^{r}\frac{dr}{1-\frac{\mathcal{B}(r)}{r}}=\pm\int_{r_0}^{r}\frac{dr}{1-\frac{4M^2\zeta^4}{r^6}}.
\end{equation}
In terms of this, the wormhole throat is at $l(r_0)=0$ and the two signs correspond to the two regions connected by the throat. In the present case, $1-\frac{\mathcal{B}(r)}{r}=\mu(r)=1-\frac{4M^2\zeta^4}{r^6}$, as discussed before.


\section{Constraints on $\zeta$ from Observations}
\label{sec:4}
\begin{figure}[h]
\centering
\begin{subfigure}{.5\textwidth}
\centering
\includegraphics[scale=0.8]{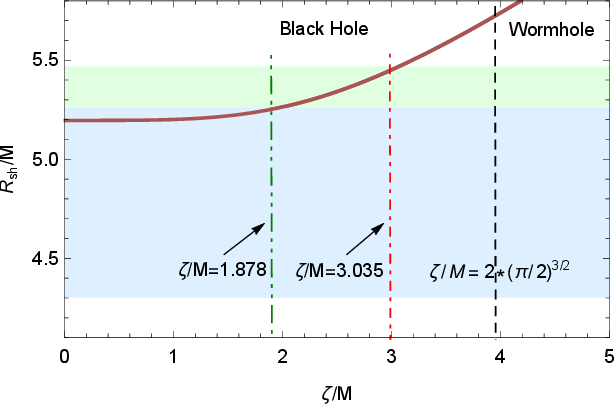}
\subcaption{SgrA*}
\label{fig:Shplt1}
\end{subfigure}%
\begin{subfigure}{.5\textwidth}
\centering
\includegraphics[scale=0.8]{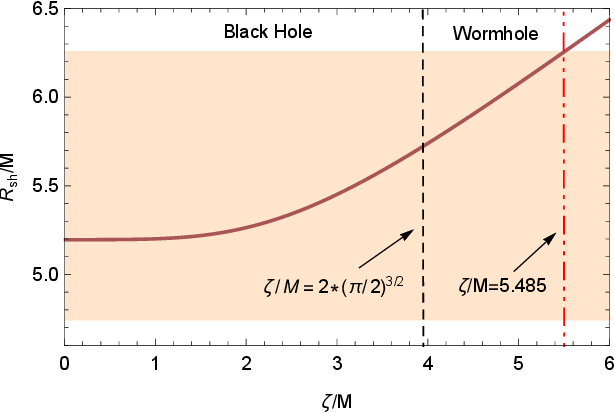}
\subcaption{M87*}
\label{fig:Shplt2}
\end{subfigure}
\caption{Plots of $R_{\text{sh}}$ (in units of $M$) as a function of $\zeta$ (in units of $M$). The shaded regions represent the observational ranges of $R_{\text{sh}}/M$ for (a) SgrA* and (b) M87*. The values of $\zeta/M$ that fall within these shaded regions should be allowed. In (a), the `light blue' shaded region represents VLTI and `light green' region corresponds to Keck observations. Constraints on $\zeta$ have been shown with vertical lines: green dot-dashed for VLTI and red dot-dashed for Keck. In (b), the `light orange' shaded region represents M87* observational range. Constraints on $\zeta$ have been shown with vertical red dot-dashed line. The black-dashed vertical lines in both the plots correspond to the extremal value of $\zeta$.}
\label{fig:zeta-bound}
\end{figure}

In this and the following sections, we have considered $q=0$ case only, for convenience. Here, we put bounds on the quantum parameter $\zeta$ based on the observational results by the EHT Collaboration of shadows of the central objects of the M87* and SgrA* galaxies. A recent study on the shadows and light rings for this spacetime can be found in \cite{Liu3}. A discussion on the possible bounds on $\zeta$ are also provided there. However, since this aspect is crucial for our discussion, we shall briefly elaborate upon it for completeness.\\
Light rays coming from a distant source and having impact parameter less than the critical impact parameter corresponding to the photon sphere (i.e., $b<b_m$) do not comprise any turning point. They continue to spiral towards the center and never reach the observer. Therefore, such rays produce a dark shade in the observer's sky, forming a `shadow', whose edge is traced out by the rays having $b=b_m$. The spacetime under consideration being spherically symmetric, the corresponding shadow will be circular and its radius with respect to the observer for $q=0$ is evaluated from Eq. (\ref{eq:bm}) as
\begin{equation}
R_{\text{sh}}=b_m=\left[\frac{1}{r_m^2}-\frac{1}{\zeta^2}\arcsin\left(\frac{2M\zeta^2}{r_m^3}\right)\right]^{-1/2}.
\end{equation}
Recent observation by the EHT Collaboration for M87* galaxy provide a mass of $M=(6.5\pm0.7)\times10^9M_{\odot}$, a distance of $D=(16.8\pm0.8)$ Mpc from the observer and the angular diameter of the shadow to be $\Delta\theta_{\text{sh}}=42\pm3$ $\mu$as \cite{EHTM1}, where $\Delta\theta_{\text{sh}}=2R_{\text{sh}}/D$. The observed value of the shadow diameter $d_{\text{sh}}(=2R_{\text{sh}})$, in dimensionless unit, is found to be \cite{EHTM1}
\begin{equation}
\frac{d_{\text{sh}}}{M}=11.0\pm1.5.
\end{equation}
Therefore, from the M87* observation, we obtain the following range of $R_{\text{sh}}$:
\begin{equation}
4.75M\leq R_{\text{sh}}\leq 6.25M.
\label{eq:boundM87}
\end{equation}
On the other hand, observations on the SgrA* have reported a fractional deviation parameter $\delta$, which measures the fractional deviation of the observed shadow diameter from that of the Schwarzschild one, given by
\begin{equation}
\delta=\frac{d_{\text{sh}}}{d_{\text{sh,Sch}}}-1=\frac{R_{\text{sh}}}{R_{\text{sh,Sch}}}-1=\frac{R_{\text{sh}}}{3\sqrt3M}-1 \implies R_{\text{sh}}=3\sqrt{3}(\delta+1)M
\label{eq:frac-dev}
\end{equation}
The EHT collaboration has provided two sets of values of $\delta$ from the Keck and the Very Large Telescope Interferometer (VLTI) observations, as given below \cite{EHTS1,EHTS4,EHTS6}
\begin{equation}
\delta=\begin{cases}
-0.04^{+0.09}_{-0.10} & \text{(Keck)}\\
-0.08^{+0.09}_{-0.09} & \text{(VLTI)}.
\end{cases}
\end{equation}
Therefore, the values of $\delta$ lie in the range $-0.14\leq\delta\leq+0.05$ (Keck) and $-0.17\leq\delta\leq+0.01$ (VLTI). From Eq. (\ref{eq:frac-dev}), the corresponding ranges of the shadow radius of SgrA* can be evaluated as,
\begin{equation}
4.47M\leq R_{\text{sh}}\leq5.46M ~~~ \text{(Keck)} ~~~~~ \text{and} ~~~~~ 4.31M\leq R_{\text{sh}}\leq5.25M ~~~ \text{(VLTI)}.
\label{eq:boundSgr}
\end{equation}
Plots of $R_{\text{sh}}$ (units of $M$) as functions of $\zeta$ (in units of $M$), along with allowed ranges from observations, are shown in Fig. (\ref{fig:Shplt1}) for SgrA* and in Fig. (\ref{fig:Shplt2}) for M87*.
The shaded regions in both the plots represent the observational ranges of $R_{\text{sh}}/M$. The deep brown curve represents the variation of $R_{\text{sh}}/M$ as a function of $\zeta/M$. The black-dashed vertical line in both the plots in Fig. (\ref{fig:zeta-bound}) correspond to $\zeta/M=2\left(\frac{\pi}{2}\right)^{3/2}\approx3.937$. The left side of this line ,i.e., $\zeta/M<2\left(\frac{\pi}{2}\right)^{3/2}$ represents the black hole and the right side of it stands for the wormhole. In case of M87* observations, it can be seen from Fig. (\ref{fig:Shplt2}) that the maximum allowed value of $\zeta$ is $5.485M$ (denoted by the red dot-dashed vertical line). So the observational constraint on $\zeta$ from M87* observations turns out be $\zeta/M\in[0,5.485]$. Therefore, both the black hole and the wormhole cases are allowed in this scenario.\\
On the other hand, in case of SgrA* observations shown in Fig. (\ref{fig:Shplt1}), the `light blue' shaded region represents the observational range of $\zeta/M$ for VLTI and `light green' shaded region corresponds to the Keck observation. In case of VLTI, the maximum allowed value of $\zeta$ is found out to be $1.878M$ (denoted by the green dot-dashed vertical line) and in case of Keck, it is $3.035M$ (denoted by the red dot-dashed vertical line). So, the observational constraints form SgrA* results come out to be $\zeta/M\in[0,1.878]$ for VLTI and $\zeta/M\in[0,3.035]$ for Keck. Since both the maximum allowed values of $\zeta$ are less than $2M\left(\frac{\pi}{2}\right)^{3/2}\approx3.937M$, they fall within the black hole case only. Hence, M87* observations support both black hole and wormhole cases; whereas SgrA* observations put more stringent constraints supporting black hole case only. We shall now explore the strong lensing features of this spacetime considering both black hole and wormhole cases in the following sections.


\section{Lensing of Light in the Strong Deflection Limit}
\label{sec:5}
\begin{figure}[h]
\centering
\begin{subfigure}{.5\textwidth}
\centering
\includegraphics[scale=0.9]{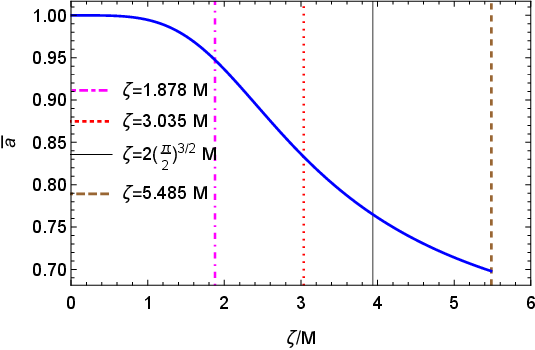}
\subcaption{Variation of $\bar{a}$ with $\zeta/M$}
\label{fig:abar-plot}
\end{subfigure}%
\begin{subfigure}{.5\textwidth}
\centering
\includegraphics[scale=0.9]{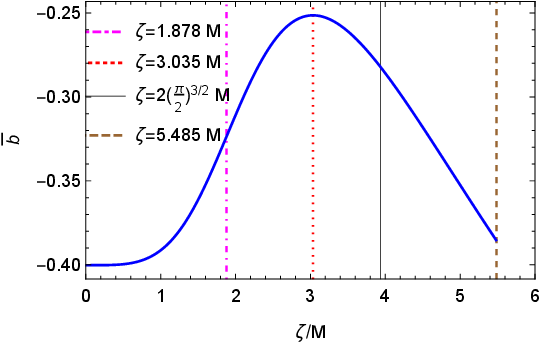}
\subcaption{Variation of $\bar{b}$ with $\zeta/M$}
\label{fig:bbar-plot}
\end{subfigure}
\caption{Plots of (a) $\bar{a}$ and (b) $\bar{b}$ as functions of $\zeta$ (in units of $M$) for $q=0$. Markers are provided with vertical lines for $\zeta=1.878M$ (VLTI), $\zeta=3.035M$ (Keck), $\zeta=2\left(\frac{\pi}{2}\right)^{3/2}M$ and $\zeta=5.485M$ (M87*).}
\label{fig:abar-bbar}
\end{figure}

For convenience, let us begin our study of strong lensing with a general static, spherically symmetric spacetime represented by the following line element
\begin{equation}
ds^2=-F(r)dt^2+G(r)dr^2+C(r)\left(d\theta^2+\sin^2\theta d\phi^2\right).
\label{eq:genstatic}
\end{equation}
Comparing Eq. (\ref{eq:genstatic}) with Eq. (\ref{eq:genmetric}) for $q=0$, we get $F(r)=A^0(r)$, $G(r)=\mu(r)^{-1}[A^0(r)]^{-1}$ and $C(r)=r^2$. Here, we first write down the relevant expressions of strong lensing in terms of the general metric functions ($F(r)$, $G(r)$) of Eq. (\ref{eq:genstatic}) and at the end, replace $F(r)$, $G(r)$ and their derivatives with the expressions of $A^0(r)$ and $\mu(r)$, to obtain the final results in terms the metric functions of Eq. (\ref{eq:genmetric}) with $q=0$.\\
A ray of light having impact parameter $b>b_m$, where $b_m$ refers to the critical impact parameter corresponding to a photon sphere, starts from a distant source, moves towards the central lensing object, takes a turn at $r_{\text{tp}}$ (where $\dot{r}|_{r_\text{tp}}=0$) and finally escapes to a distant observer. The bending angle $\alpha(r_{\text{tp}})$ for such a light ray can be obtained as \cite{Bozza02,Tsu17}
\begin{equation}
\alpha(r_{\text{tp}})=I(r_{\text{tp}})-\pi,
\label{eq:deflection1}
\end{equation}
where
\begin{equation}
I(r_{\text{tp}})=2\int^{\infty}_{r_{\text{tp}}}\frac{dr}{\sqrt{\frac{R(r)C(r)}{G(r)}}}~,~
R(r)= \left(\frac{F(r_{\text{tp}})C(r)}{F(r)C(r_{\text{tp}})}-1 \right).
\label{eq:I1}
\end{equation}
Again, it has been discussed earlier that the impact parameter ($b$) for such a ray having a turning point at $r_{\text{tp}}$ is given as, $b(r_{\text{tp}})=\sqrt{C(r_{\text{tp}})/F(r_{\text{tp}})}$. Eliminating $r_{\text{tp}}$ from $\alpha(r_{\text{tp}})$ and $b(r_{\text{tp}})$, the bending angle can be expressed as a function of the impact parameter, i.e., $\alpha(b)$. In a spacetime having photon sphere ($r_m$) only, the bending angle diverges logarithmically as $r_{\text{tp}}\to r_m$ or $b\to b_m$ when the light ray approaches the photon sphere from $b > b_m$ side. A comprehensive study of strong lensing due to a photon sphere can be found in \cite{Bozza02,Tsu17}. The corresponding divergence of bending angle due to photon sphere (for both black holes and wormholes) in the strong deflection limit, i.e.,  $r_{\text{tp}}\to r_{m}$ or $b\to b_m$, is given as \cite{Tsu17} 
\begin{equation}
\alpha(b)=-\bar{a}\log \left( \frac{b}{b_m}-1 \right) +\bar{b} +\mathcal{O}((b-b_m)\log(b-b_m)),
\label{eq:strong_alpha_1}
\end{equation}
where $\bar{a}$ and $\bar{b}$ are given by
\begin{equation}
\bar{a}=\sqrt{\frac{2G(r)F(r)}{C^{''}(r)F(r)-C(r)F^{''}(r)}}\Biggr\rvert_{r_m}~,~
\bar{b}=\bar{a}\log \left[r^{2}\left(\frac{C^{''}(r)}{C(r)}-\frac{F^{''}(r)}{F(r)}\right)\right]\Biggr\rvert_{r_m} +I_{R}(r_{m})-\pi.
\label{eq:strong_bbar_1}
\end{equation}
Here, prime denotes a derivative with respect to $r$, and $I_{R}(r_{m})$ is a constant \cite{Tsu17}. Replacing $F(r)$, $G(r)$ and their derivatives with $A^0(r)$ and $\mu(r)$ in the above equation, we can obtain the final forms of $\bar{a}$ and $\bar{b}$ for the spacetime under consideration.
Fig. (\ref{fig:abar-bbar}) depicts the variations of these two coefficients as functions of $\zeta/M$ for $q=0$. Markers are provided with vertical lines for $\zeta=1.878M$ (VLTI), $\zeta=3.035M$ (Keck), $\zeta=2\left(\frac{\pi}{2}\right)^{3/2}M$ and $\zeta=5.485M$ (M87*). We can see that $\bar{a}$ decreases trivially with increasing $\zeta$; whereas $\bar{b}$ varies in a non-trivial way. In case of the other two types of black hole solutions of EQG \cite{Zhang1}, both the strong field coefficients decrease with increasing the quantum parameter, as shown in Ref. \cite{Liu1}.
The specific values of $\bar{a}$ and $\bar{b}$ are also provided in Table (\ref{Table1}) for the above set of values of $\zeta$.
\begin{table}[h]
\centering
\caption{Values of $\bar{a}$ and $\bar{b}$ for different values of $\zeta$ (in units of $M$). The second column is for $\zeta=0.0$ corresponding to the Schwarzschild black hole. The third and fourth columns are for $\zeta=1.878M$ and $\zeta=3.035M$ respectively, which represent the maximum allowable values of $\zeta$ constrained from the VLTI and Keck observations of SgrA* respectively. The fifth column corresponds to the extremal black hole for $\zeta=2\left(\frac{\pi}{2}\right)^{3/2}M$. And the last column stands for $\zeta=5.485M$ which represents the maximum allowable value of $\zeta$ from M87* observation. The last column represents the wormhole case, and rest of the others represent black hole cases. Numerical values are obtained in natural units.}
\begin{tabular}{| c | c | c | c | c | c |} 
\hline\hline
& $\zeta=0$ & $~\zeta=1.878M~$ & $~\zeta=3.035M~$ & $~\zeta=2\left(\frac{\pi}{2}\right)^{3/2}M~$ & $~\zeta=5.485M~$ \\
& (Schwarzschild) & (VLTI) & (Keck) & (Extremal) & (M87*) \\
\hline
$~\bar{a}~$ & 1 & 0.9473 & 0.8330 & 0.7649 & 0.6979 \\
$~\bar{b}~$ & -0.40023 & -0.3238 & -0.2514 & -0.2822 & -0.3856 \\
\hline\hline
\end{tabular}
\label{Table1}
\end{table}\\
Plots of bending angle ($\alpha$) versus impact parameter ($b$) (in units of $M$) are shown in Fig. (\ref{fig:bending}) for the set of values of $\zeta$ as considered in Table (\ref{Table1}), i.e., $\zeta=\{0.0,~ 1.878M,~ 3.035M,~ 2\left(\frac{\pi}{2}\right)^{3/2}M,~ 5.485M\}$. Here, plots are obtained from the exact expression of $\alpha$ using Eqs. (\ref{eq:deflection1}) and (\ref{eq:I1}). It can be seen from the figure that the Schwarzschild and VLTI plots almost merge with each other. So, if $\zeta$ is constrained according to the VLTI observation, it will be difficult to distinguish this black hole from the Schwarzschild one, making it a candidate for Schwarzschild black hole mimicker. With increase of $\zeta$, the deviations from the Schwarzschild black hole become larger for the Keck and extremal black hole cases. On the other hand, if $\zeta$ is constrained according to the M87* observations, the deviations for the wormhole cases are found out to be significant from the Schwarzschild black hole.
\begin{figure}[h]
\centering
\includegraphics[scale=0.75]{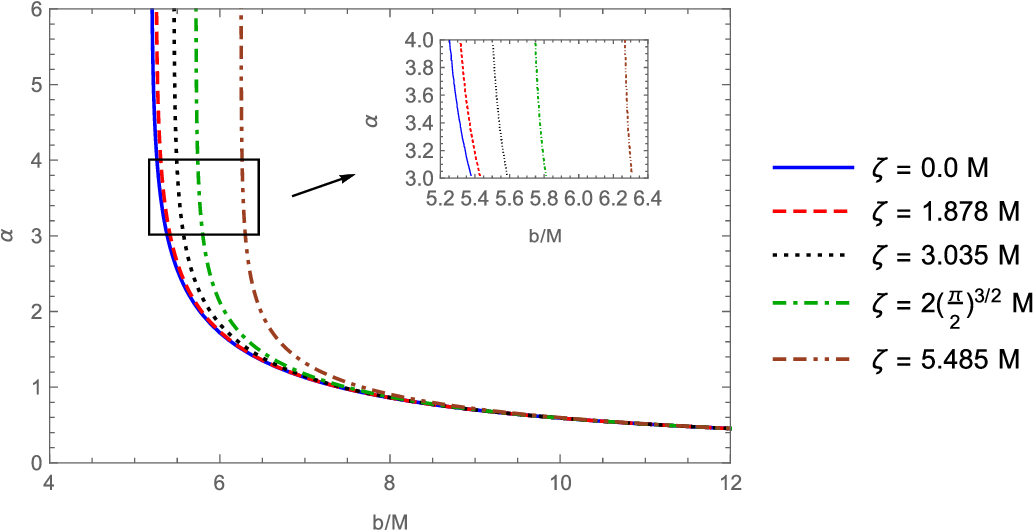}
\caption{Plots of bending angle $\alpha$ versus impact parameter $b$ (in units of $M$) for $\zeta=\{0.0,~ 1.878M,~ 3.035M,~ 2\left(\frac{\pi}{2}\right)^{3/2}M,~ 5.485M\}$.}
\end{figure}
\label{fig:bending}


\section{Strong Lensing Observables}
\label{sec:6}
In this section, we shall discuss about various observables corresponding to the relativistic images formed due to photon sphere in the context of strong gravitational lensing of the quantum gravity spacetime under consideration. Expressions of most of these observables for the relativistic images formed just outside the photon sphere have been obtained in Ref. \cite{Bozza02}. The lens equation in the strong deflection limit, considering the source, lensing object and observer to be nearly aligned (thin lens approximation), is given as \cite{Bozza01}
\begin{equation}
\delta=\theta-\frac{D_{LS}}{D_{OS}}\Delta\alpha_n,
\end{equation}
where $\delta$ is the angular separation between the light source and the lensing object, $\theta$ is the angular separation between the image and the lensing object, $D_{LS}$ is the distance between the lensing object and the light source, $D_{OS}$ is the distance between the observer and the light source. Since light rays take a number of turns (say, $n$) before reaching the observer in the strong field limit, the corresponding offset of deflection angle is represented by $\Delta\alpha_n=\alpha(\theta)-2\pi n$. Note that $D_{OS}=D_{OL}+D_{LS}$, where $D_{OL}$ is the distance between the observer and the lensing object.

The angular position of $n$-th relativistic image formed just outside the photon sphere can be approximated by \cite{Bozza01, Bozza02}
\begin{equation}
\theta_n\approx\theta_n^0+\frac{b_m e_n (\delta-\theta_n^0)D_{OS}}{\bar{a}D_{LS}D_{OL}}, \quad e_n=e^{\frac{\bar{b}-2n\pi}{\bar{a}}},
\end{equation}
where $\theta_n^0=\frac{b_m}{D_{OL}}(1+e_n)$ is the image position corresponding to $\alpha\left(\theta_n^0\right)=2\pi n$. It is to be noted that the second term in the above equation is negligible as compared to $\theta_n^0$. Therefore, for practical purposes, we can roughly take
\begin{equation}
\theta_n=\theta_n^0=\frac{b_m}{D_{OL}}(1+e_n)=\theta_{\infty}(1+e_n),
\end{equation}
where $\theta_{\infty}=b_m/D_{OL}$ is the angular position of the relativistic image formed at the photon sphere. Moreover, the magnification of the $n$-th relativistic image is given as \cite{Bozza02}
\begin{equation}
\mu_n=\left(\frac{\delta}{\theta}\frac{\partial\delta}{\partial\theta}\right)^{-1}\bigg\rvert_{\theta^0_n}=\frac{b_m^2 e_n(1+e_n)D_{OS}}{\bar{a}\delta D_{OL}^2 D_{LS}}.
\end{equation}
It should be noted that $\theta_1$ corresponding to $n=1$ represents the angular position of the outermost relativistic image just outside the photon sphere, and $\theta_{\infty}$ corresponds to the innermost one at the photon sphere. This implies that the angular positions of the relativistic images decrease with increasing $n$. It is usually observed that the inner images are closely packed together with the innermost one; leaving only the first (outermost) one ($n=1$) to be resolved from the rest. Therefore, two more lensing observables can be defined, namely $s_1$ representing the angular separation between first and the rest (effectively the innermost one), and $r_1$ denoting the relative magnification of the outermost image, defined by the ratio of the magnification of the first image and the sum of magnifications of the rest. They are given by \cite{Bozza02}
\begin{equation}
s_1=\theta_1-\theta_\infty=\theta_\infty e_1,
\end{equation}
\begin{equation}
r_1=\frac{\mu_1}{\sum\limits_{m=2}^\infty \mu_m}\approx e^{\frac{2\pi}{\bar{a}}}.
\label{eq:outer_flux_ratio}
\end{equation}\\
Moreover, since light rays forming relativistic images in the strong deflection limit may take turn around the lensing object multiple times before coming to the observer, travel time of photons through different light paths may be significantly different for different relativistic images. The difference of travel time for different images may produce detectable effects, giving rise to another important lensing observable, namely Time Delay. Substantive discussion on this observable can be found in Ref. (\cite{Bozzadelay}). In case of a general static, spherically symmetric spacetime represented by the line element in Eq. (\ref{eq:genstatic}), the time delay ($\Delta T$) between $n$-th and $m$-th relativistic images is given as \cite{Bozzadelay}
\begin{equation}
\Delta T^s_{n,m} =2\pi(n-m)b_m+2\sqrt{\frac{G_m}{F_m}}\sqrt{\frac{b_m}{p}}~e^{\frac{\bar{b}}{2\bar{a}}}\left(e^{-\frac{2m\pi\mp\delta}{2\bar{a}}}-e^{-\frac{2n\pi\mp\delta}{2\bar{a}}}\right),
\end{equation}
when the two images are on the same side of the optic axis (line joining lens and observer) with the upper negative sign before $\delta$ signifying both the images and the source (all three together) on the same side of the optic axis, and the lower positive sign before $\delta$ signifying both the images together on the same side of the optic axis and the source on the opposite side of the optic axis, whereas
\begin{equation}
\Delta T^o_{n,m} =[2\pi(n-m)-2\delta]b_m+2\sqrt{\frac{G_m}{F_m}}\sqrt{\frac{b_m}{p}}~e^{\frac{\bar{b}}{2\bar{a}}}\left(e^{-\frac{2m\pi-\delta}{2\bar{a}}}-e^{-\frac{2n\pi+\delta}{2\bar{a}}}\right),
\end{equation}
when the two images are on the opposite side of the optic axis, i.e., one of the image and the source are on one side together and the other image is on the opposite side of the optic axis. Here, $p=\frac{C_m''F_m-C_mF_m''}{4\sqrt{F_m^3C_m}}$. When the source is almost aligned with the optic axis ($\delta\ll2\pi$), it can be shown that $\Delta T^o_{n,n}\ll\Delta T^s_{n,m\neq n}\approx\Delta T^o_{n,m\neq n}$ \cite{Bozzadelay}. Moreover, the second terms on the RHS of the above equations are much smaller than the first terms ($\sim 1\%$ contribution). Therefore, considering thin lens approximation, the expression for $\Delta T$ considered for the present analysis is
\begin{equation}
\Delta T_{n,m}=2\pi(n-m)b_m=2\pi(n-m)\theta_{\infty}D_{OL}, \quad [m\neq n].
\end{equation}
Upon describing the necessary formalism, we are now in a position to analyze the observables in the context of the effective gravity spacetime under consideration, considering both the SgrA* and M87* observations. For the present analysis, we have restored $G$ and $c$ to replace $M$ by $\frac{GM}{c^2}$ where in the last expression $M$ represents the actual mass obtained from observation. The observer-lens distance ($M_{OL}$) has also been taken from observational data. Moreover, we have set $M_{LS}=M_{OL}=\frac{1}{2}M_{OS}$.\\
In case of SgrA*, the values we have considered are, $M=4.0\times10^6M_{\odot}$ and $D_{OL}=8.35$ Kpc \cite{EHTS1,EHTS4,EHTS6}. Whereas, for M87*, the corresponding values are, $M=6.5\times10^9M_{\odot}$ and $D_{OL}=16.8$ Mpc \cite{EHTM1}. In addition, we have considered $\delta=5^\circ$ for all the computations. The results have been depicted through figures and tabular forms, considering both SgrA* and M87* observations.
\begin{figure}[t]
\centering
\begin{subfigure}{.5\textwidth}
\centering
\includegraphics[scale=0.8]{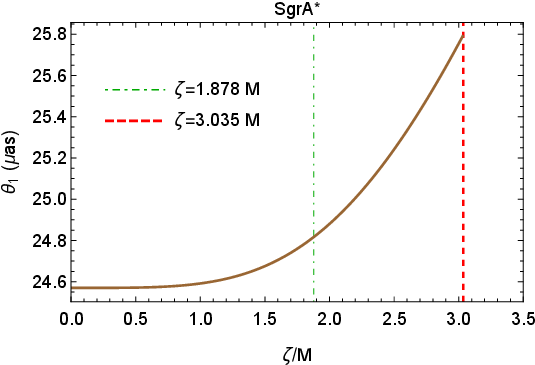}
\end{subfigure}%
\begin{subfigure}{.5\textwidth}
\centering
\includegraphics[scale=0.76]{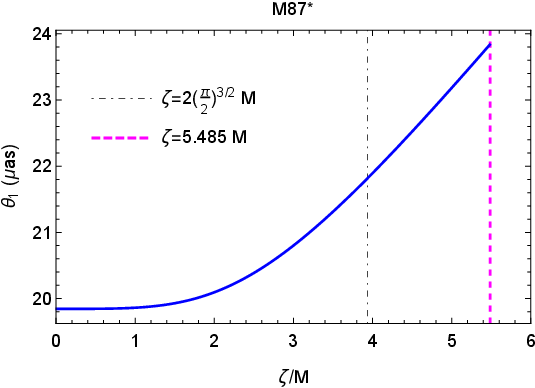}
\end{subfigure}
\\ [2ex]
\begin{subfigure}{.5\textwidth}
\centering
\includegraphics[scale=0.8]{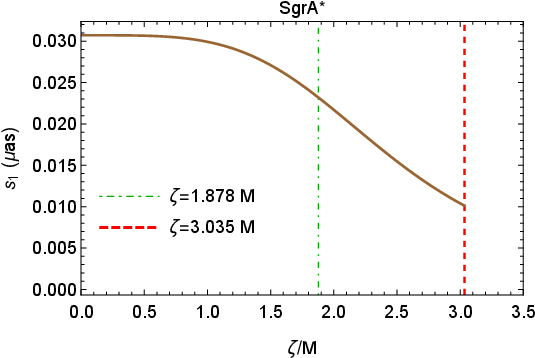}
\end{subfigure}%
\begin{subfigure}{.5\textwidth}
\centering
\includegraphics[scale=0.8]{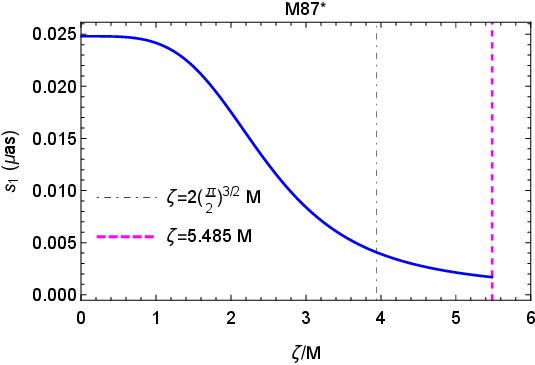}
\end{subfigure}
\caption{Plots of $\theta_1$ (top panel) and $s_1$ (bottom panel) as functions of $\zeta/M$. All angles are in `micro-arcsecond'. The left panel corresponds to SgrA* and the right panel represents M87* observations. In left panel, markers with vertical lines are provided signifying VLTI (green, dot-dashed line at $\zeta=1.878M$) and Keck (red, dashed line at $\zeta=3.035M$) observations. In right panel, black dot-dashed vertical line at $\zeta=2\left(\frac{\pi}{2}\right)^{3/2}M$ represents the transition between black hole and wormhole cases, and the magenta dashed line at $\zeta=5.485M$ stands for the maximum allowed limit of $\zeta$ for wormhole case from M87* observation.}
\label{fig:th1-s1}
\end{figure}

In Fig. ($\ref{fig:th1-s1}$), the angular position ($\theta_1$) of the first (outermost) image just outside the photon sphere (top panel) and the separation ($s_1=\theta_1-\theta_{\infty}$) of it from the innermost image at the photon sphere (bottom panel) are plotted as functions of $\zeta/M$\footnote{Note that, in all the figures, $\zeta/M$ is taken instead of the actual values of $\zeta$ along the $x$-axis. This is because the values of $\zeta$ are obtained by multiplying the numerical values of $\zeta/M$ with the observational values of $M$, which just change the numerical scaling. It does not produce any change in the nature of the plots. So, $\zeta/M$ itself is retained for convenience.}. The left panel corresponds to the SgrA* and the right panel represents the M87* observations. It can be seen from the plots that $\theta_1$ rises with increasing values of $\zeta$, whereas $s_1$ decreases. This clearly signifies that it will become increasingly difficult to resolve the first image from the rest with higher values of $\zeta$. Importantly, relativistic images for the wormhole (region in the right side of the black-dashed vertical line in the right panel of M87*) turn out to be more difficult to resolve than the black hole counterpart.
\begin{figure}[h]
\centering
\begin{subfigure}{.5\textwidth}
\centering
\includegraphics[scale=0.76]{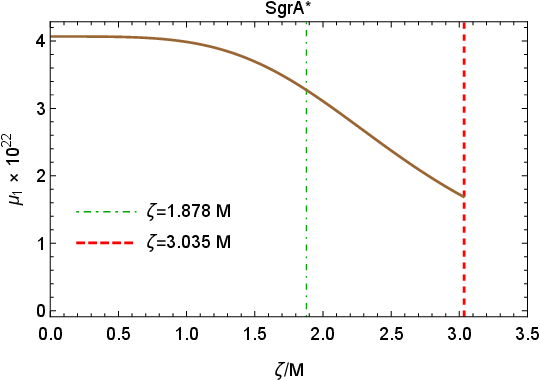}
\end{subfigure}%
\begin{subfigure}{.5\textwidth}
\centering
\includegraphics[scale=0.8]{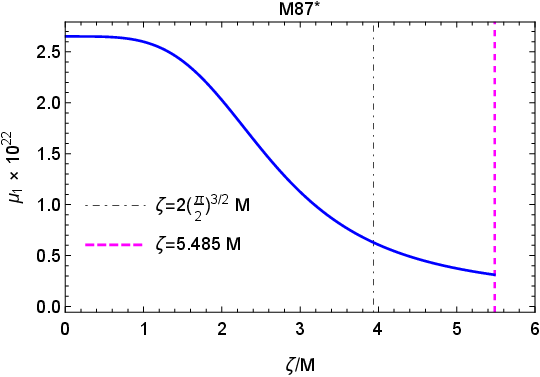}
\end{subfigure}
\\ [2ex]
\begin{subfigure}{.5\textwidth}
\centering
\includegraphics[scale=0.8]{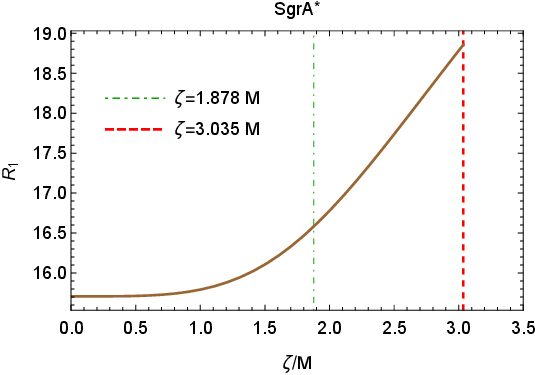}
\end{subfigure}%
\begin{subfigure}{.5\textwidth}
\centering
\includegraphics[scale=0.76]{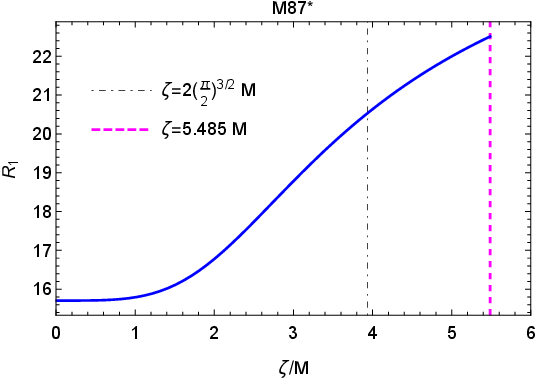}
\end{subfigure}
\caption{Plots of $\mu_1$ (top panel) and $\mathcal{R}_1$ (bottom panel) as functions of $\zeta/M$. The left panel corresponds to SgrA* and the right panel represents M87* observations. The markers with vertical lines carry the same meaning as described in the caption of Fig. (\ref{fig:th1-s1}).}
\label{fig:mu1-R1}
\end{figure}
Similarly, in Fig. (\ref{fig:mu1-R1}), the absolute magnification $\mu_1$ and the relative flux, converted to magnitude using $\mathcal{R}_1=2.5\log r_1$, of the first (outermost) image are plotted against $\zeta/M$. It can be seen that the images get increasingly de-magnified with increase of $\zeta$, while the relative flux $\mathcal{R}_1$ increases. Here also, relativistic images for wormhole will have much lower magnification as compared to black hole. Nature of variation of these observables with respect to the quantum parameter ($\zeta$) are similar to those of the other two black hole solutions of EQG \cite{Liu1}.
\begin{figure}[h]
\centering
\begin{subfigure}{.5\textwidth}
\centering
\includegraphics[scale=0.83]{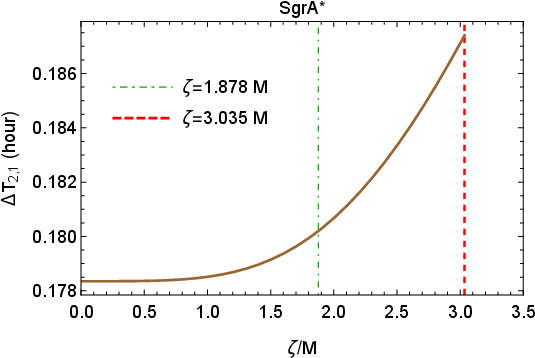}
\end{subfigure}%
\begin{subfigure}{.5\textwidth}
\centering
\includegraphics[scale=0.8]{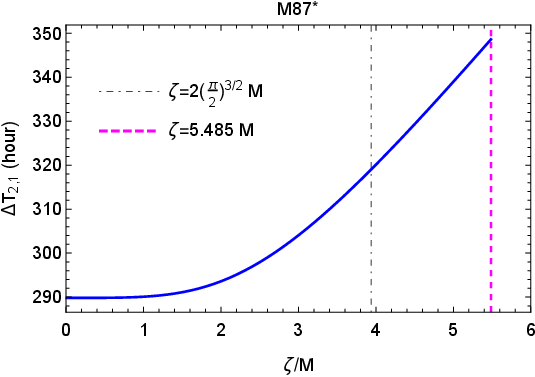}
\end{subfigure}
\caption{Plots of time delay $\Delta T_{2,1}$ (in hour) as functions of $\zeta/M$. The left panel corresponds to SgrA* and the right panel represents M87* observations. The markers with vertical lines carry the same meaning as described in the caption of Fig. (\ref{fig:th1-s1}).}
\label{fig:T21}
\end{figure}

In Fig. (\ref{fig:T21}), the time delay between the first and second relativistic images ($\Delta T_{2,1}$) is plotted against $\zeta/M$. It is evident from the figure that the time delay comes out to be of the orders of minutes for SgrA* and around several days for M87*. Such short time delays in SgrA* provide little chance of observation, whereas it yields measurable time delay in case of M87* observation. Interestingly, effects of $\zeta$ on time delay comes out to be quite different as compared to those of the first two types of black hole solutions of EQG \cite{Zhang1}. In one of the solution, time delay decreases with increase of $\zeta$, while for the other, it remains constant \cite{Liu1}. Here, for the third type, time delay increases with $\zeta$. While most of the other observables provide negligible opportunity for detection and distinction between black hole and wormhole cases, time delay may become crucial in this aspect, at least for ultra compact objects having masses of the order of M87*. Specific values of these lensing observables have also been presented in Table (\ref{tab:Tab-obs}) for convenience.
\begin{table}[h!]
\centering
\caption{Specific values of lensing observables for both SgrA* and M87*. All angles are in `micro-arcsecond' and $r_n$ is converted to magnitude using $\mathcal{R}_n=2.5\log r_n$. Here, we have considered $M=4.0\times 10^6 M_{\odot}$, $D_{OL}=D_{LS}=8.35$ Kpc for SgrA* and $M=6.5\times 10^9 M_{\odot}$, $D_{OL}=D_{LS}=16.8$ Mpc for M87*, and $\delta=5^\circ$ for both. The second, third and fourth columns correspond to the SgrA* black hole cases and the last column represents the M87* wormhole case. Since the transitional case (fifth column) falls outside the domain of SgrA* observation, its values are obtained using the M87* data.}
\begin{tabular}{| c | c | c | c | c | c |}
\hline\hline
& $\zeta=0$ & $\zeta=1.878M$ & $\zeta=3.035M$ & $\zeta=2\left(\frac{\pi}{2}\right)^{3/2}M$ & $\zeta=5.485M$ \\
& (Schwarzschild) & (VLTI) & (Keck) & (Extremal) & (M87*) \\
\hline
$\theta_1$ ($\mu$as) & 24.5701 & 24.8167 & 25.7955 & 21.8191 & 23.8410 \\
$\theta_{\infty}$ ($\mu$as) & 24.5394 & 24.7935 & 25.7854 & 21.8150 & 23.8393 \\
$s_1$ ($\mu$as) & 0.0307 & 0.0232 & 0.0101 & 0.0041 & 0.0017 \\
$\mu_1\times 10^{22}$ & 4.0647 & 3.2725 & 1.6861 & 0.6276 & 0.3105 \\
$\mathcal{R}_1$ & 15.7064 & 16.5811 & 18.8558 & 20.5355 & 22.5085 \\
$\Delta T_{2,1}$ (hour) & 0.1784 & 0.1802 & 0.1874 & 319.001 & 348.602 \\
\hline\hline
\end{tabular}
\label{tab:Tab-obs}
\end{table}

 
\section{Conclusion}
\label{sec:7}
In this work, we study gravitational lensing in the strong deflection limit by the third type of EQG solution proposed in Ref. \cite{Zhang2}. The line element of the spacetime depends on two free parameters $q$ and $\zeta$. It is shown that the spacetime, for $q=0$, is asymptotically flat for any finite value of $\zeta$, and reduces to the Schwarzschild black hole in the limit $\zeta\to0$. Whereas, for even integer values of $q$, the spacetime asymptotically resembles to the Schwarzschild-de Sitter (SdS) geometry and for odd integer values of $q$, it produces effects similar to that of a negative mass object \cite{Zhang2}. This article concentrates on $q=0$ case only, to study its strong lensing features. We observe that, for a specific range of values of $\zeta\in[0,2\left(\frac{\pi}{2}\right)^{3/2}M]$, the spacetime contains event horizon. Whereas beyond this range, there exists a minimum value of the radial coordinate which is shown to act like a wormhole throat. Hence, the spacetime contains both black hole and wormhole solutions with $\zeta=2\left(\frac{\pi}{2}\right)^{3/2}M\approx$ being the transitional or extremal case. Using EHT data from SgrA* and M87* observations, the allowable ranges of $\zeta$ are obtained as $\zeta\in[0,1.878M]$ (VLTI), $\zeta\in[0,3.035M]$ (Keck), and $\zeta\in[0,5.485M]$ (M87*). It signifies that the wormhole case is ruled out by SgrA* observations, while both black hole and wormhole cases are possible by M87* observations.

We, then, obtain the deflection angle parameters in the strong field regime following \cite{Bozza02,Tsu17}, generated their plots as functions of $\zeta$ and compared them with the Schwarzschild case. In addition, we analyze the strong lensing observables and described the effects of the quantum parameter $\zeta$ on them through their representative plots. It is found that, with increase of $\zeta$, it becomes difficult to resolve and detect the relativistic images outside the photon sphere. However, time delay between the first two relativistic images comes out to be of the order of several days in case of M87* observation, which may have the potential for detection in future observations.

To conclude, our theoretical study suggests that the free quantum parameters play an important role in the strong lensing characteristics and observables which may be used as an astrophysical test of EQG. The present analysis are also important in the sense that it opens up multiple exciting research prospects such as studying the strong lensing features of this spacetime when $q\neq 0$, analyzing the physics of accretion disks in this background and generating its image etc. We intend to envisage them in future.


\section{Acknowledgments}
The author would like to thank P. Banerjee for poring over the manuscript, providing with valuable suggestions and engaging in meaningful discussions. The author also thanks the anonymous reviewer for providing critical comments that improve the presentation and contents of the manuscript.


\begin{appendices}
\section{Brief review of the effective canonical framework: from action to the metric}
\label{appen}
In the classical framework of Hamiltonian formulation of GR, the 4-D manifold ($\mathcal{M}$, $g_{\rho\sigma}$), adapted to coordinates $x^\rho$, is decomposed into a ($3+1$)-D foliation by noninteracting spacelike hypersurfaces ($\Sigma$, $h_{ab}$), adapted to coordinates $y^a$, a lapse scalar ($N$) and a shift vector ($N^a$). The corresponding decomposed line element is given by,
\begin{equation}
ds^2=g_{\rho\sigma}dx^{\rho}dx^{\sigma}=-N^2dt^2+h_{ab}\left(dx^a+N^a dt\right)\left(dx^b+N^b dt\right)
\end{equation}
Starting with the 4-D Einstein-Hilbert action for pure gravity in bulk, its decomposition into standard 3+1 form is given as \cite{Poisson,Jha}
\begin{equation}
S_G = \frac{1}{16\pi} \int_{t_1}^{t_2}dt \int_{\Sigma} \left({}^{(3)}R+K_{ab}K^{ab}-K^2\right) N\sqrt{h}~d^3y,
\end{equation}
where ${}^{(3)}R$ is the Ricci scalar associated with the induced metric $h_{ab}$, $K_{ab}$ is the extrinsic curvature of $\Sigma$ embedded in $\mathcal{M}$, $K=h^{ab}K_{ab}$ and $h=h^{ab}h_{ab}$. From the above action or the associated Lagrangian density, extracting the Hamiltonian density is standard in the literature, and the resulting action in terms of Hamiltonian density is given as, \cite{Poisson,Jha}
\begin{equation}
S_G = \int_{t_1}^{t_2}dt \int_{\Sigma}\left[p^{ab}\dot{h}_{ab} - (N\mathcal{H}+N^a\mathcal{H}_a)\right]~d^3y,
\end{equation}
where $p^{ab}$ is the canonical momenta conjugate to $h_{ab}$, $\mathcal{H}$ and $\mathcal{H}_a$ are respectively the Hamiltonian and diffeomorphism (momentum) constraint densities over $\Sigma$. The constant factor $\frac{1}{16\pi}$ is absorbed into the integrand. It should be noted that the action admits only $h_{ab}$ and $p_{ab}$ as the dynamical degrees of freedom; whereas $N$ and $N^a$ act as Lagrange multipliers.

In Ref. \cite{Zhang2}, the authors consider spherical symmetry for which the spacelike 3-manifold is further decomposed as $\Sigma=\mathcal{X}\times\mathbb{S}^2$, adapted to coordinates $y^a\equiv (r,\theta,\phi)$, where $\mathcal{X}$ denotes a 1-D manifold and $\mathbb{S}^2$ represents the 2-sphere. The phase space dynamical variables $(h_{ab},p_{ab})$ or $(h_{ab},K_{ab})$ are reduced to two pairs of canonical dynamical variables $(K_I, E^I)$ ($I=1,2$), defined on the quotient manifold $\mathcal{X}$, representing extrinsic curvature and densitized triad for both the radial and angular sectors. As canonical pairs, $(K_I, E^I)$ satisfy the fundamental Poisson bracket relations. Integrating the 4-D action over the angular coordinates (employing spherical symmetry) yields the following symmetry-reduced gravitational action governing the dynamics,
\begin{equation}
\tilde{S}_G = \int_{t_1}^{t_2}dt \int \left[E^1\dot{K}_1 + 2E^2\dot{K}_2 - (N\tilde{\mathcal{H}}+N^r\tilde{\mathcal{H}}_r)\right]dr,
\end{equation}
where $\tilde{\mathcal{H}}$, $\tilde{\mathcal{H}}_r$ are the corresponding 1-D constraint densities defined on $\mathcal{X}$. Due to spherical symmetry, the two other twisting components of $\mathcal{H}_a$ must vanish, and the lapse \& shifts must be functions of $r$ only. The constant $4\pi$ of angular integration is absorbed into the integrand. The above action can be varied with respect to the canonical fields $(K_I, E^I)$ as well as the Lagrange multipliers $(N, N^a)$. Varying the action with respect to $(K_I, E^I)$ gives rise to the Hamilton's equations for the fields,
\begin{equation}
\dot{K}_I=\{K_I, H[N]+H_r[N^r]\}, \quad\quad \dot{E}^I=\{E^I, H[N]+H_r[N^r]\},
\end{equation}
where $H[N]=\int N(r)\tilde{\mathcal{H}}(r) dr$ and $H_r[N^r]=\int N^r(r)\tilde{\mathcal{H}}_r(r) dr$. Again, varying $\tilde{S}_G$ with respect to $(N, N^a)$ leads to the primary constraint equations, 
\begin{equation}
\tilde{\mathcal{H}}\approx0, \quad\quad \tilde{\mathcal{H}}_r\approx0.
\end{equation}
The weak equality `$\approx$' simply means that it needs to be satisfied over $\mathcal{X}$ only, not in between the hypersurfaces. Moreover, the constraints form a closed Poisson algebra necessary to ensure general covariance of the theory \cite{Zhang2} ,
\begin{equation}
\begin{split}
\{H_r[N_1^r],H_r[N_2^r]\}=H_r[N_1^r\partial_rN_2^r-N_2^r\partial_rN_1^r],\\
\{H[N],H_r[M^r]\}=-H[M^r\partial_rN], \quad\quad\quad\\
\{H[N_1],H[N_2]\}=H_r[S(N_1\partial_rN_2-N_2\partial_rN_1)],
\end{split}
\label{eq:H-poissn}
\end{equation}
where $S=\frac{E^1}{(E^2)^2}$ is known as the structure function which is related to the $(r,r)$ component of the inverse spatial metric.
For given lapse and shift functions, not all solutions to the Hamilton's equations represent physical states. Solutions corresponding to physical states must also satisfy the constraint equations. Therefore, the physical states lie in the constraint surface consisting of $(K_I, E^I)$ points that satisfy the constraint equations. Now, for every moment $t_0$, these $(K_I(t_0), E^I(t_0))$ fields corresponding to a physical state, the lapse $N(t_0)$ and the shift $N^r(t_0)$ can be pushed forward to the $t=t_0$ slice of $\mathcal{M}$, leading to the following metric in $\mathcal{M}$,
\begin{equation}
ds^2=g_{\rho\sigma}dx^{\rho}dx^{\sigma}=-N^2dt^2+\frac{(E^2)^2}{E^1}\left(dr+N^rdt\right)^2+E^1d\Omega^2,
\label{eq:H-met}
\end{equation}
where $d\Omega^2=d\theta^2+\sin^2\theta d\phi^2$ is the line element on $\mathbb{S}^2$. Evidently, the metric depends on the choice of $N$ and $N^r$. However, metrics obtained from different choices of $N$ and $N^r$ are related by 4-D diffeomorphism transformations in $\mathcal{M}$, ensured by the closed Poisson algebra of the constraints. Hence, the totally constraint Hamiltonian formulation of classical GR with the constraint algebra given by Eq. (\ref{eq:H-poissn}) is built generally covariant with respect to the above metric (\ref{eq:H-met}).

To incorporate quantum effects, the authors in Ref. \cite{Zhang2} adopt a semiclassical approach where the diffeomorphism constrain density ($\tilde{\mathcal{H}}_r$) retains its classical form, while the Hamiltonian constraint density ($\tilde{\mathcal{H}}$) is replaced by an effective counterpart ($\tilde{\mathcal{H}}_{\text{eff}}$) incorporating the desired quantum modifications. Accordingly, the action, Hamilton's equations and constraint equations are modified by replacing $\tilde{\mathcal{H}}$ with $\tilde{\mathcal{H}}_{\text{eff}}$ in their respective expressions. Moreover, the corresponding constraint algebra is assumed to be modified as \cite{Zhang2}
\begin{equation}
\begin{split}
\{H_r[N_1^r],H_r[N_2^r]\}=H_r[N_1^r\partial_rN_2^r-N_2^r\partial_rN_1^r],\\
\{H_{\text{eff}}[N],H_r[M^r]\}=-H_{\text{eff}}[M^r\partial_rN], \quad\quad\quad\\
\{H_{\text{eff}}[N_1],H_{\text{eff}}[N_2]\}=H_r[\mu S(N_1\partial_rN_2-N_2\partial_rN_1)],
\end{split}
\label{eq:Heff-poissn}
\end{equation}
where $\mu$ is the modification scalar function of the field variables $(K_I,E^I)$ capturing QG effects and $H_{\text{eff}}[N]=\int N(r)\tilde{\mathcal{H}}_{\text{eff}}(r) dr$. This modification breaks the general covariance of the effective theory with respect to the metric of Eq. (\ref{eq:H-met}), motivating to consider a modified metric as given below
\begin{equation}
ds^2=g^{(\mu)}_{\rho\sigma}dx^{\rho}dx^{\sigma}=-N^2dt^2+\frac{(E^2)^2}{\mu E^1}\left(dr+N^rdt\right)^2+E^1d\Omega^2. 
\label{eq:Heff-met}
\end{equation}
Note that the new structure function $\mu S$ appears in the $(r,r)$ component of the inverse spatial metric, similar to the classical case. Now, to restore general covariance of the effective theory with respect to the modified metric, a set of necessary and sufficient conditions are derived in Ref. \cite{Zhang2}. On the basis
of these conditions, a set of covariance equations satisfied by an effective mass function $M_{\text{eff}}(E^1,E^2,K_2)$ are obtained. Subsequently, $\tilde{\mathcal{H}}_{\text{eff}}$ is expressed as a function of the derivatives of $M_{\text{eff}}$. As a result, once $M_{\text{eff}}$ is determined, $\tilde{\mathcal{H}}_{\text{eff}}$ can easily be obtained from their relation.

Accordingly, a suitable effective mass function $(M_{\text{eff}})$ is proposed in Ref. \cite{Zhang2} (in the paper, it is denoted by the symbol $M^{(3)}_{\text{eff}}$). Then putting $M_{\text{eff}}$ into the covariance equations, the functional form the modification function $\mu$ is obtained. So, at this stage, we have the expressions of $M_{\text{eff}}(E^1,E^2,K_2)$, $\mu(E^1,E^2,K_2)$ and $\tilde{\mathcal{H}}_{\text{eff}}(E^1,E^2,K_1,K_2)$.
To extract the metric, the following set of gauge fixing conditions are then considered:\\
(i) the areal gauge $\Rightarrow$ $E^1(r)=r^2$, and\\
(ii) from the vanishing of $\tilde{\mathcal{H}}_r$ $\Rightarrow$ $K_1(r)=\frac{E^2(r)K_2'(r)}{r}$, \\
where `prime' denotes a derivative with respect to $r$. Using these gauge fixing conditions, $\tilde{\mathcal{H}}_{\text{eff}}$ becomes \cite{Zhang2},
\begin{equation}
\tilde{\mathcal{H}}_{\text{eff}}(r)=-\frac{E^2(r)}{r}\partial_r M_{\text{eff}}(r).
\end{equation}
where
\begin{equation}
M_{\text{eff}}(r)=h(r^2)+\frac{r\mathcal{F}(r^2)\sin[\lambda(r^2)F(r)]}{2\lambda(r^2)},
\label{eq:Meff}
\end{equation}
with $F(r)=1+K_2(r)^2-\frac{r^2}{[E^2(r)]^2}+\psi(r^2)$. Here $h(r),\mathcal{F}(r),\lambda(r)$ and $\psi(r)$ are arbitrary functions appeared as integration constants. The constraint equation $\tilde{\mathcal{H}}_{\text{eff}}(r)=0$ becomes $M_{\text{eff}}(r)=$ constant ($=M$, say), from which the following expression is obtained, using Eq. (\ref{eq:Meff}),
\begin{equation}
\sin[\lambda(r^2)F(r)]=\frac{2\lambda(r^2)}{r\mathcal{F}(r^2)}[M-h(r^2)].
\label{eq:sin-fn}
\end{equation}
Now, to get a stationary metric, the time evolution of $E^I$ $(I=1,2)$ is set to zero, i.e., $\dot{E}^I=\{E^I,H_{\text{eff}}[N]+H_r[N^r]\}=0$, $(I=1,2)$, which yields \cite{Zhang2}
\begin{equation}
N(r)=\frac{\sqrt{E^1}}{E^2(r)}=\frac{r}{E^2(r)} \quad\quad \text{and} \quad\quad \frac{N^r(r)}{N(r)}=-\cos[\lambda(r^2)F(r)]K_2(r)\mathcal{F}(r^2).
\end{equation}
Again, the metric is expressed in static Schwarzschild-like coordinates by setting $K_2(r)=0$ so that the shift, $N^r(r)=0$. Therefore, Eq. (\ref{eq:sin-fn}) simplifies to
\begin{equation}
\frac{r^2}{[E^2(r)]^2}=N(r)^2=1-\frac{q\pi+(-1)^q \arcsin\left[\frac{2\lambda(r^2)(M-h(r^2))}{r\mathcal{F}(r^2)}\right]}{\lambda(r^2)}+\psi(r^2),
\end{equation}
where $q$ is an integer. In addition, the modification function $\mu(r)$ also becomes,
\begin{equation}
\mu(r)=\mathcal{F}(r^2)^2-\frac{4\lambda(r^2)^2}{r^2}\left[M-h(r^2)\right]^2.
\end{equation}
Finally, the remaining arbitrary parameters are chosen as (see Ref. \cite{Zhang2}), $\lambda(r^2)=\frac{\zeta^2}{r^2}$, $\mathcal{F}(r^2)=1$, $h(r^2)=0$ \& $\psi(r^2)=0$, and the resulting static spherically-symmetric metric in Schwarzschild-like coordinates takes the form,
\begin{equation}
ds^{2}=g^{(\mu)}_{\rho\sigma}dx^{\rho}dx^{\sigma}=-A^{(q)}(r)dt^{2}+\frac{dr^{2}}{\mu(r)A^{(q)}(r)}+r^2(d\theta^{2}+\sin^{2}\theta d\phi^{2}),
\end{equation}
where
\begin{equation}
A^{(q)}(r)=1-(-1)^q\frac{r^2}{\zeta^2}\arcsin\left(\frac{2M\zeta^2}{r^3}\right)-\frac{q\pi r^2}{\zeta^2}, ~~~ \mu(r)=1-\frac{4M^2\zeta^4}{r^6}.
\end{equation}

\end{appendices}


\end{document}